\documentclass[useAMS,usenatbib]{mn2e}
\usepackage{graphicx,natbib,amsmath}

\begin{document}
			
\title[Adaptive pupil masking]{Adaptive pupil masking for quasi-static speckle suppression}

\author[J. Osborn]{James Osborn\thanks{E-mail:
josborn@ing.puc.cl}\\
Centre for Astro-Engineering, Pontificia Universidad Catolica de Chile, Vicu\~{n}a Mackenna 4860, Santiago, Chile}	

\maketitle

\begin{abstract}			
Quasi-static speckles are a current limitation to faint companion imaging of bright stars. Here we show through simulation and theory that an adaptive pupil mask can be used to reduce these speckles and increase the visibility of faint companions. This is achieved by placing an adaptive mask in the conjugate pupil plane of the telescope. The mask consists of a number of independently controllable elements which can either allow the light in the subaperture to pass or block it. This actively changes the shape of the telescope pupil and hence the diffraction pattern in the focal plane. By randomly blocking subapertures we force the quasi-static speckles to become dynamic. The long exposure PSF is then smooth, absent of quasi-static speckles. However, as the PSF will now contain a larger halo due to the blocking, the signal to noise ratio ({\it SNR}) is reduced requiring longer exposure times to detect the companion. For example, in the specific case of a faint companion at $5\lambda/D$ the exposure time to achieve the same {\it SNR} will be increased by a factor of 1.35. In addition, we show that the visibility of companions can be greatly enhanced in comparison to long-exposures, when the dark speckle method is applied to short exposure images taken with the adaptive pupil mask. We show that the contrast ratio between PSF peak and the halo is then increased by a factor of approximately 100 (5 magnitudes), and we detect companions 11 magnitudes fainter than the star at 5$\lambda/D$ and up to 18 magnitudes fainter at 22.5$\lambda/D$.
\end{abstract}
\begin{keywords}
instrumentation: adaptive optics -- instrumentation: high angular resolution 
\end{keywords}

\section{Introduction}

Detecting the faint reflected or self-luminous signal from extrasolar planetary companions close to a bright parent star is a technically difficult task. With the development of sophisticated image analysis and Adaptive Optics (AO) systems on several modern large (8~m class) telescopes it is now possible. AO is required to both increase the peak intensity of the point-spread function (PSF) and to concentrate the photons which are scattered into a diffuse halo by the atmosphere back into the diffraction limited core. Dedicated high-contrast imaging instruments, such as HiCAIO (Subaru, \citeauthor{Hodapp08}, 2008), SPHERE (VLT, \citeauthor{Beuzit08}, 2008) and GPI (Gemini, \citeauthor{Macintosh06}, 2006), are designed to incorporate eXtreme AO (XAO) systems and sophisticated coronagraphs to reject the light from the star whilst conserving the few photons from the angularly separated companion.  Quasi-static speckles, mimicking the signal of faint companions, are now limiting the detection capabilities of these instruments (e.g. \citeauthor{Fitzgerald06}, 2006; \citeauthor{Soummer07}, 2007). Quasi-static speckles in the focal plane are caused by non-common path errors and uncorrected aberrations in the primary mirror and other optical and mechanical components. If these aberrations were entirely deterministic they could be subtracted. However, quasi-static speckles are slowly varying aberrations making calibration difficult.

The current most popular methods for quasi-static speckle reduction are PSF subtraction techniques such as, PSF estimation (e.g. \citeauthor{Lafreniere07}, 2007)  angular differential imaging (ADI, \citeauthor{Marois06}, 2006), simultaneous spectral differential imaging (SSDI, \citeauthor{Smith87}, 1987) and, in the case of reflected light, polarimetric differential imaging \citep{Seager00}. SSDI and PDI both require certain properties from the target. ADI is more generic but if the speckles evolve during the observation the suppression provided by the technique reduces dramatically. The temporal decorrelation timescale of these quasi-static speckles is an important factor when estimating the performance of image subtraction techniques which have proved themselves to be efficient at Strehl ratios of the order of 20--40\% \citep{Martinez12}. At high Strehl ($>$80\%) the quasi-static speckles will become even more dominant as the PSF halo is reduced further. In this high Strehl regime the speckle coherence timescale is unknown. \citeauthor{Martinez12} show that the quasi-static speckles become unstable over a timescale of a few seconds on a laboratory XAO test bench. It is thought that the evolution of the speckle pattern was primarily caused by temperature fluctuations and so, on a dedicated instrument, this could be more controlled.

Several other interesting and inventive techniques are also being developed in order to further enhance the probability of detecting faint companions. \cite{Ribak08} demonstrate it is possible to enhance the contrast by placing a rotating eccentric mask in the pupil plane. This breaks the symmetry between the telescope pupil and the focal plane causing the quasi-static speckles to move, the companion will however stay fixed at the same location. \cite{Gladysz08,Gladysz09} use the statistical distribution difference of on-axis and off-axis PSF to differentiate between real sources and speckles. It is likely that only the combination of several of these techniques in conjunction with AO and a coronagraph will result in the highest contrast ratios in modern instrumentation.

Here we propose to combine the idea of breaking the symmetry of the optical system and the sensor by changing the pupil function (similar to \citeauthor{Ribak08}) with an adaptive pupil mask (APM) \citep{Osborn09}. The APM is positioned in the conjugate plane of the telescope pupil. This pupil mask consists of a number of independently controllable elements. The simplest design would be a segmented mirror where each segment can either reflect the light on-axis into the remaining optical system or off-axis into a baffle. This is also similar to the speckle decorrelation (or phase--boiling) method \citep{Saha02}. The phase--boiling method involves adding additional phase aberrations to the optical path in order to force the quasi-static speckles to be more dynamic. This was later disproven by \cite{Sivaramakrishnan02} as the original phase aberrations are still present and so the quasi-static speckles are also still present, simply hidden within a field of dynamic speckles. Here we boil the speckles by introducing amplitude aberrations instead of adding phase aberrations, an important difference as the interference pattern will no longer contain the same speckle pattern, but will actually be completely different.

By changing the shape of the pupil we modify the diffraction pattern in the pupil plane. Light that once interfered constructively to form a quasi-static speckle at a given location will now not.  However, speckles will be formed in other locations in the focal plane. The quasi-static speckles will be forced to be dynamic, removing any dependance on speckle timescale. By changing the pupil function many times during an exposure these speckles will average out into a smooth PSF, albeit with an additional halo due to diffraction through the pupil mask (Babinet's principle). This will mean that longer integration times are required for a companion to become visible with the same signal to noise ratio (SNR). However, the quasi-static speckles will be substantially reduced, reducing the complexity of the companion identification problem.

In addition to the APM reducing the static speckles into a smooth halo, we can also expose after each configuration of the mask separately. As the quasi-static speckles are now dynamic we can use other image manipulation methods to further enhance the image. Dark speckle (DS) imaging, first proposed by \cite{Labeyrie95}, is a technique designed for the detection of faint companions in the presence of dynamic speckles induced by the turbulent atmosphere. Briefly, after low order AO correction to focus the majority of the photons in to the diffraction limited core and a coronagraph to reject the photons from the star there still remains some atmospheric turbulence induced speckles in the PSF halo in short exposures. In this speckle halo, at some locations the wavefront will interfere destructively and result in a zero photon event. As the atmosphere evolves and traverses the telescope field of view, the position of these nulls will change in the focal plane. However, at the position of a faint companion the probability of a zero photon event is considerably lower. Therefore, by counting the number of times each pixel records a zero photon event in each short exposure we can generate a `dark map' where the position of the companion will have a value lower than the rest of the image. Modern XAO systems are designed with target residual wavefront error (WFE) of the order of a few nm and so there is actually very little in the way of residual atmospheric speckle. However, as the quasi-static speckles are now dynamic we can use the dark speckle method to suppress the now dynamic quasi-static speckles. 

The DS method is intended to be used to enhance AO corrected images. Each DS exposure must be short otherwise the AO residual speckles will average, reducing the speckle nulls. Here we develop the technique for quasi-static speckles, in which case the exposure time no longer needs to be short and we record the minimum value over a number of mask configurations.

An additional advantage of the pupil mask is that it is a configurable device allowing it to act in several different observing modes. In `adaptive' mode the mask can also be used to put a hard limit on the residual atmospheric WFE.  Adaptive pupil masks like this have been shown to be able to reduce the PSF halo and actually increase PSF peak intensity despite removing photons \citep{Osborn09}, making it useful in scenarios where only low-order AO or even no AO is available. The APM could be used in `static' mode as a non-redundant aperture mask (e.g. \citeauthor{Kopilovich84}, 1984) or partially-redundant aperture mask (e.g. \citeauthor{Buscher93}, 1993), used in many modern high contrast imaging instruments. If the APM is sufficiently high order it can also be used to emulate any binary shaped pupil-plane mask coronagraph (e.g. \citeauthor{Nisenson01}, 2001; \citeauthor{Kasdin03}, 2003). The configuration of these static masks can easily be changed to experiment with different positions, element sizes and configurations, a current area of active research.

In section 2 we describe the simulation, in section 3 we introduce the APM and show how it will reduce the quasi-static speckles, section 4 describes application of the dark speckle method to APM, in section 5 we discuss the results and we conclude in section 6.

\section{Simulation}
\subsection{Quasi-static speckles}
\begin{figure}
   \centering
   $\begin{array}{l@{\hspace{-3.5mm}}r}
   \includegraphics[width=44mm]{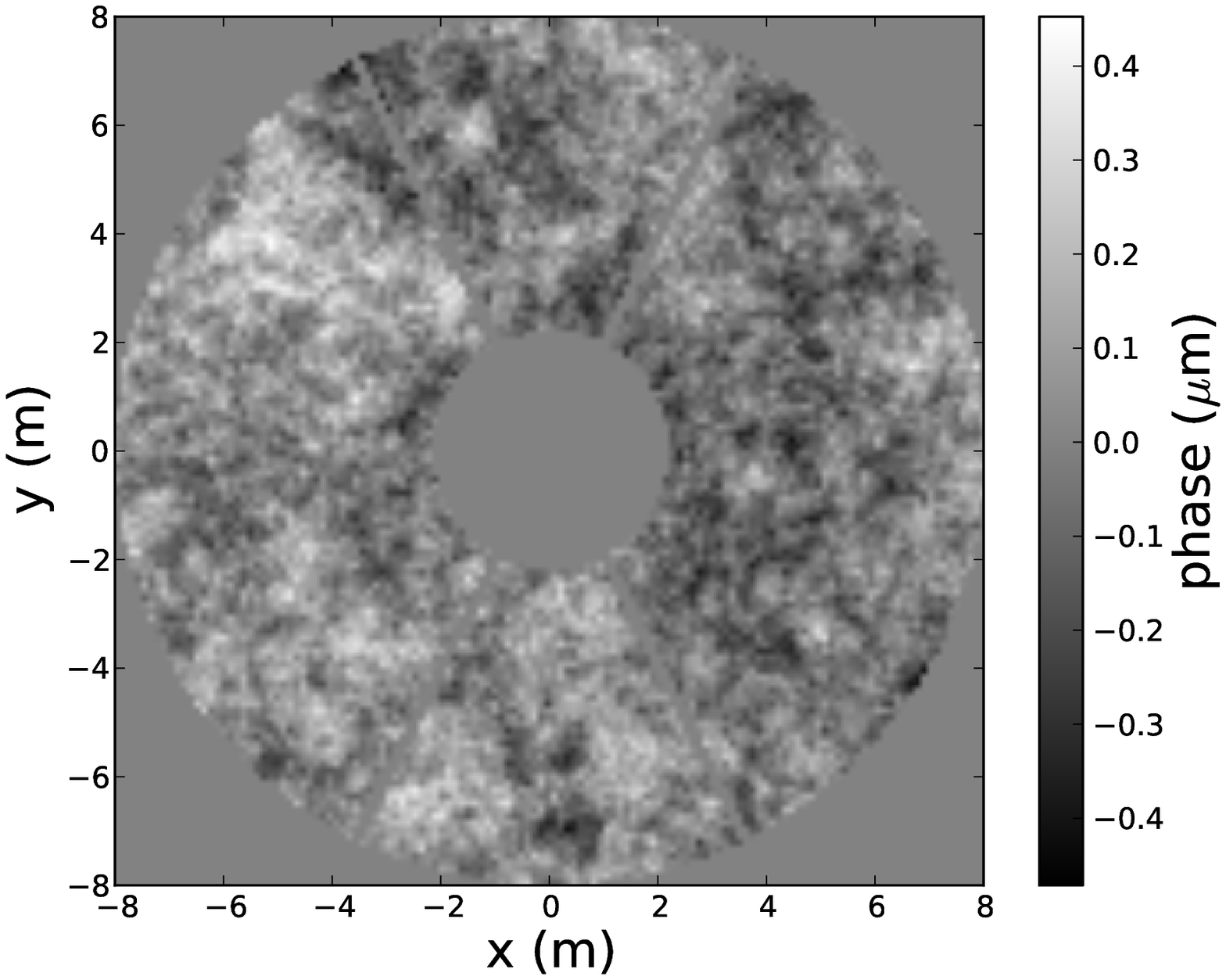} &
      \includegraphics[width=44mm]{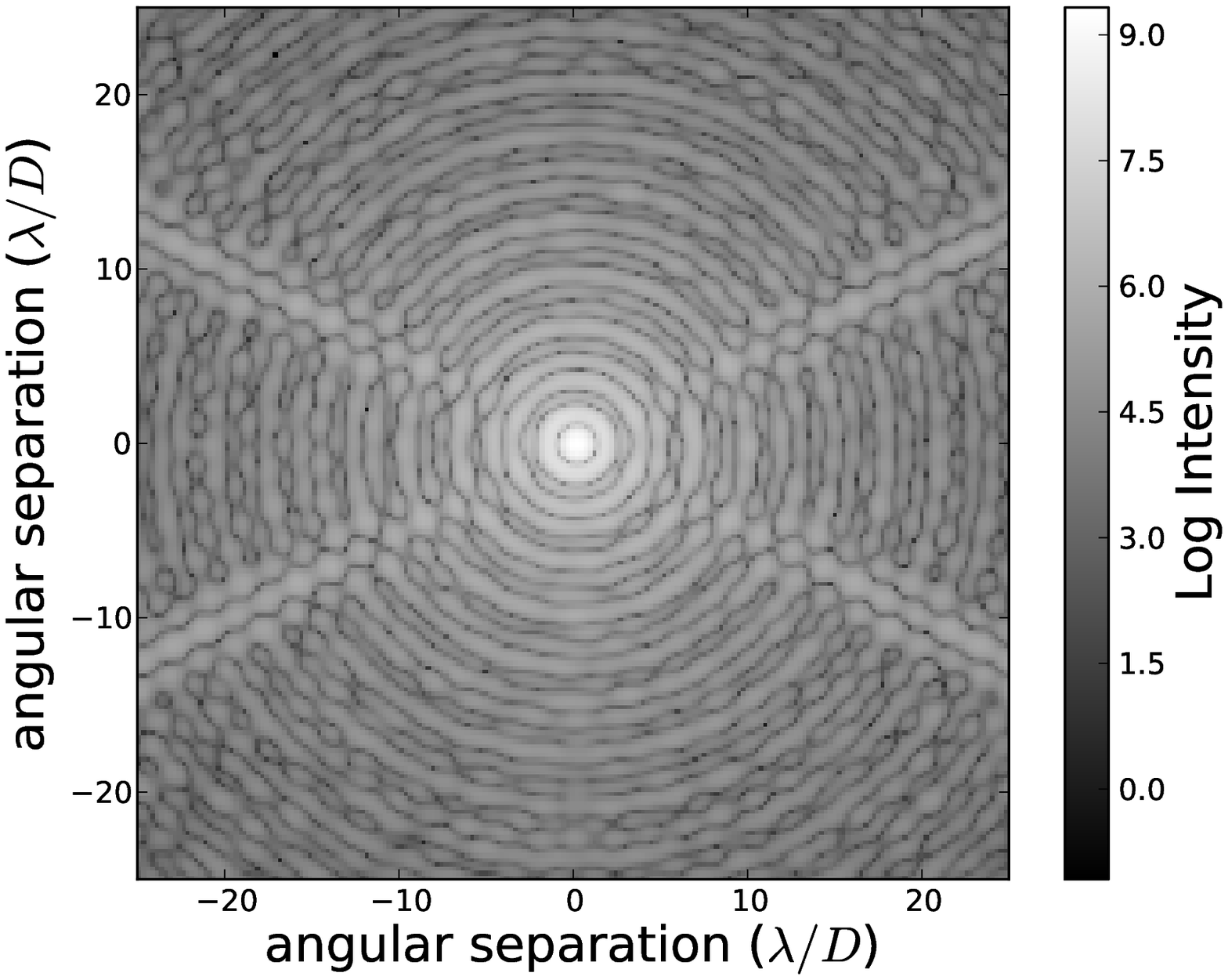} \\
   \end{array}$
    \includegraphics[width=44mm]{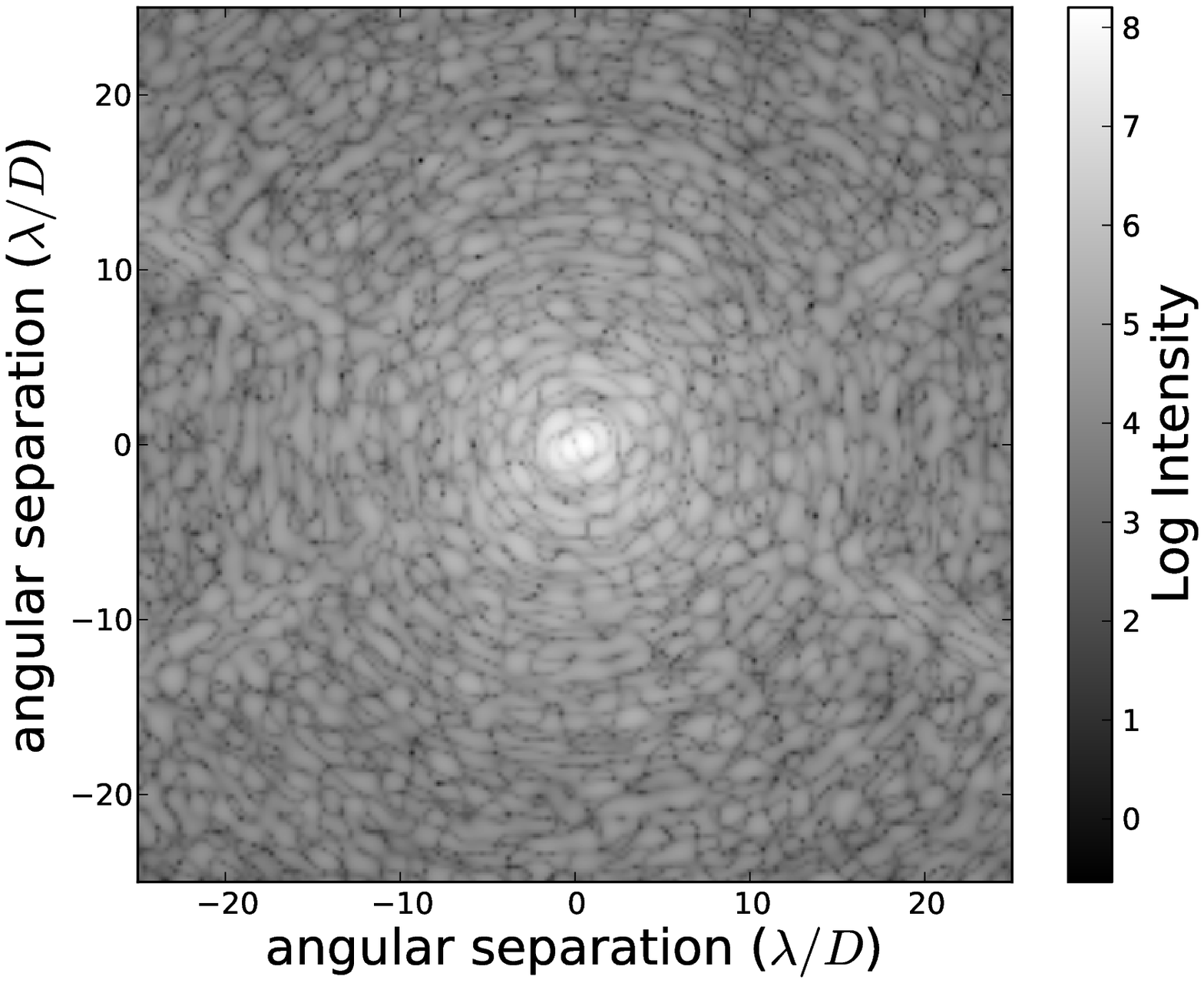}
   \caption{The primary mirror phase aberrations and pupil including telescope secondary support spiders. The RMS error is one fifth of a wavelength. On the right is a log-scaled simulated image of a star with AO residuals, photon noise, read noise, sky background noise and quasi-static speckles. The lower panel shows the speckles. Plotted is the absolute difference between two images, one with quasi-static speckles and one without.}
   \label{fig:av_pupil}
\end{figure}

The combined mirror and optics aberrations are simulated using the method of \cite{Cavarroc06} using the $\kappa^{-2}$ (where $\kappa$ is the spatial frequency) power law in the spatial power spectrum, as defined in \cite{Duparre02}. Figure~\ref{fig:av_pupil} shows a resultant static phase error, the corresponding image and an example of the speckles found by subtracting an image generated with quasi-static speckles from one without.


\subsection{Adaptive optics}

A Monte-Carlo simulation has been developed to test the concept. The simulation includes Poisson noise, sky background noise (assuming 14$^{\mathrm{th}}$ magnitude) and 10e$^{-}$ read noise. Residual adaptive optics phase aberrations are also included in the simulations. We assume a Strehl ratio of 90\% (a measure of AO residual error), consistent with predictions for XAO systems. This residual wavefront error results in a speckled intensity pattern around the diffraction limited core, over time this averages to a smooth halo. 

There are a few choices for simulating this AO corrected PSF:
\begin{itemize}
\item{We can assume a basic two part PSF, with a central diffraction limited core and a larger, in spatial extent, Gaussian halo with relative energy contributions corresponding to the desired Strehl. This would be the simplest case. However, it is a very simple approximation and would result in a smooth estimate of the infinitely long exposure PSF. This will not include any low intensity and high spatial frequency residual speckles in the image, which are important to include when developing a method to reduce such speckles.}
\item{We can use a fully analytical approach using statistical distributions to estimate the infinitely long exposure AO corrected PSF. As above, this will also result in a smooth PSF absent of residual speckles, but will be more accurate as it will include a deformable mirror model (subaperture sizes, mirror type etc.).}
\item{We can use a complete end-to-end Monte-Carlo simulation. This is more complicated and requires in depth modelling of the whole optical system (including wavefront sensors, reconstructor, mirror dynamics etc.) which are system dependant. This is a more complicated solution which is difficult to develop, test and calibrate.}
\item{The last option is a combination of the two points above. We use a statistical expression to generate AO corrected phase screens - removing the need for wavefront sensors, reconstructor and deformable mirror dynamics, but retaining the non-deterministic nature of the PSF. This will be an approximation which is not reflected by any real system but it will give an estimate of the AO corrected PSF containing residual PSF speckles averaged over whatever time scale we desire.}
\end{itemize}
Due to the instrument independent nature of this work we only need an approximation to an AO corrected PSF with the desired wavefront error. For this reason the corrected PSF is generated using the last method in the list above.

The AO corrected PSF is estimated by summing over a number of independently realised instantaneous PSFs. Each of which is generated by a semi-analytical method of spatially filtering a von Karman phase screen with an AO transfer function and Fourier transforming to form an image.

An AO system will reduce the spatial power spectrum at low spatial frequencies, the effect of which can be modelled by a high pass filter, $H(\kappa d/2)$ \citep{Greenwood78},
\begin{equation}
	H(\kappa d/2) = 1 - \left( \frac{2J_{1}\left(\kappa d / 2 \right)}{\kappa d / 2} \right)^{2}  - 16\left(2/ \kappa d \right)^{2} J_{2}^{2}\left(\kappa d / 2 \right),
\end{equation} 
where $d$ is the diameter of the subapertures and $J_{n}$ is a Bessel function of the first kind of order $n$. The equation given is for a segmented mirror with tip/tilt and piston correction. This filter function only includes the deformable mirror fitting. Wavefront sensor errors (noise, non-linearity etc.), deformable mirror dynamics, latency and reconstructor errors are not considered.

The AO residual phase spectral density, $\Phi_{\rm AO}$ is then given by,
\begin{equation}
\Phi_{\rm AO}(\kappa) =H(\kappa d/2) \Phi_{\rm atmos}(\kappa)
\label{eqn:AOpowerspect}
\end{equation}
where $\Phi_{\rm atmos}(\kappa)$ is the von Karman spatial phase power spectrum ($\Phi_{\rm atmos}(\kappa)=r_0^{-5/3}\left( \kappa^2 + 1/L_0^2\right)^{-11/6}$, $r_0$ is the Fried parameter - a measure of the strength of the turbulence and $L_0$ is the outer scale of the atmospheric turbulence).

The AO corrected phase screen is then given by \citep{Ellerbroek2002},
\begin{equation}
\phi_\mathrm{AO} = \frac{0.1517}{\sqrt{2}}\left(\frac{W}{r_0}\right)^{5/6} \Re\mathcal{F}\left[ \sqrt{\Phi_{\rm AO}(\kappa)}(\chi(\kappa)+i\chi^{\prime} (\kappa))\right],
\label{eq:PS}
\end{equation}
where $W$ is the width of the phase screen and $(\chi(\kappa)+i\chi^{\prime} (\kappa))$ is a randomly generated repeatable white noise field.

The filtered optical turbulence phase, $\phi_\mathrm{AO}$, is added to the static phase aberrations, $\phi_{\mathrm{static}}$. This assumes that the two are independent. In fact, the AO system will attempt to correct some of the static aberrations and so $\phi_{\mathrm{static}}$ is also spatially filtered by the AO filter function. Therefore, $\Phi_{\mathrm{AO, static}} = H(\kappa d/2)\Phi_ {\mathrm{static}}$, can be used to replace $\Phi_{\rm AO}(\kappa)$ in equation~\ref{eq:PS} to generate the quasi static phase aberrations. The PSF is given by,
\begin{equation}
I(\rho, \theta)= \sum_0^N{ |\mathcal{F}\left[ {P(\xi) \exp{(-i(\phi_{\mathrm{static}}(\xi, \eta) + \phi_{\mathrm{AO}}(\xi, \eta,t)} }\right]|^{2} },
\end{equation}
where $\mathcal{F}$ is the fourier transform operator, $P(\xi)$ is the pupil function and is equal to 1 when $\xi < D/2$ and 0 otherwise, $D$ is the diameter of the telescope aperture, $\rho$ and $\theta$ are the polar co-ordinates in the focal plane, $\xi$ and $\eta$ are the pupil plane variables in polar co-ordinate space, $N$ is the number of simulation iterations and $t$ denotes the time variable.


\subsection{Simulation parameters}
The process is repeated and a long exposure image is built up for as many iterations as required. The telescope diameter is 8~m, $d=0.18$~m, $r_0=0.12$~m, $L_0=30$~m, the observing wavelength is 1.6~$\mu$m with a bandwidth of 0.23~$\mu$m (H-band), the exposure time is 30~s and the pixel scale is $0.25\lambda/D$.

No coronagraph is included in the simulations, this is because modern coronagraphic techniques are as numerous as they are complicated. This means that any achieved contrast ratios can not be directly compared to those from any high contrast imaging instruments. Instead we compare to a standard image with AO but without any other form of image manipulation.

Figure~\ref{fig:av_pupil} (top right) shows the simulated focal plane image of a bright target and nine fainter companions. The companions are distributed from the centre to the bottom of the image. The central star has a magnitude of 4 and the companions have magnitudes 14-- 22, reducing by one magnitude for each step ($2.5\lambda/D$) outwards.


\section{Adaptive pupil masking}
A re-configurable mask is used to block the light of a chosen fraction, $f$, of the pupil. This is done with a spatial light modulator with $44\times44$ elements. In each iteration a random selection of these elements are flipped to block an area of the pupil. The pupil function therefore changes in each iteration. As the image can be thought of as the interference pattern of the pupil function multiplied by the complex amplitude of any phase aberrations, the image in the focal plane will therefore also be modified. The random nature of the APM is important as this means that it is completely independent from any other optical component, it is modular. No additional information or optical components (e.g. wavefront sensors) are required.

The image is now given by,
\begin{multline}
I(\rho, \theta)= \sum_0^N |\mathcal{F} \left[ M(\xi,\eta,t)P(\xi) \right. \\
            \left. \exp{(-i (\phi_{\mathrm{static}}(\xi, \eta) + \phi_{\mathrm{AO}}(\xi, \eta, t) ))}    \right]|^{2},
\end{multline}
where $M(\xi, \eta, t)$ is the time dependant mask function. In the simulation presented here, we block a random 15\% of the telescope aperture in each iteration. The actual fraction can be chosen and optimised by the user depending on system specific parameters. Here we choose 15\% as this is consistent with the throughput of many coronagraphic systems \citep{Ribak08}.

\begin{figure}
   \centering
$\begin{array}{l@{\hspace{-5mm}}c@{\hspace{-4mm}}r}
  \includegraphics[width=30mm]{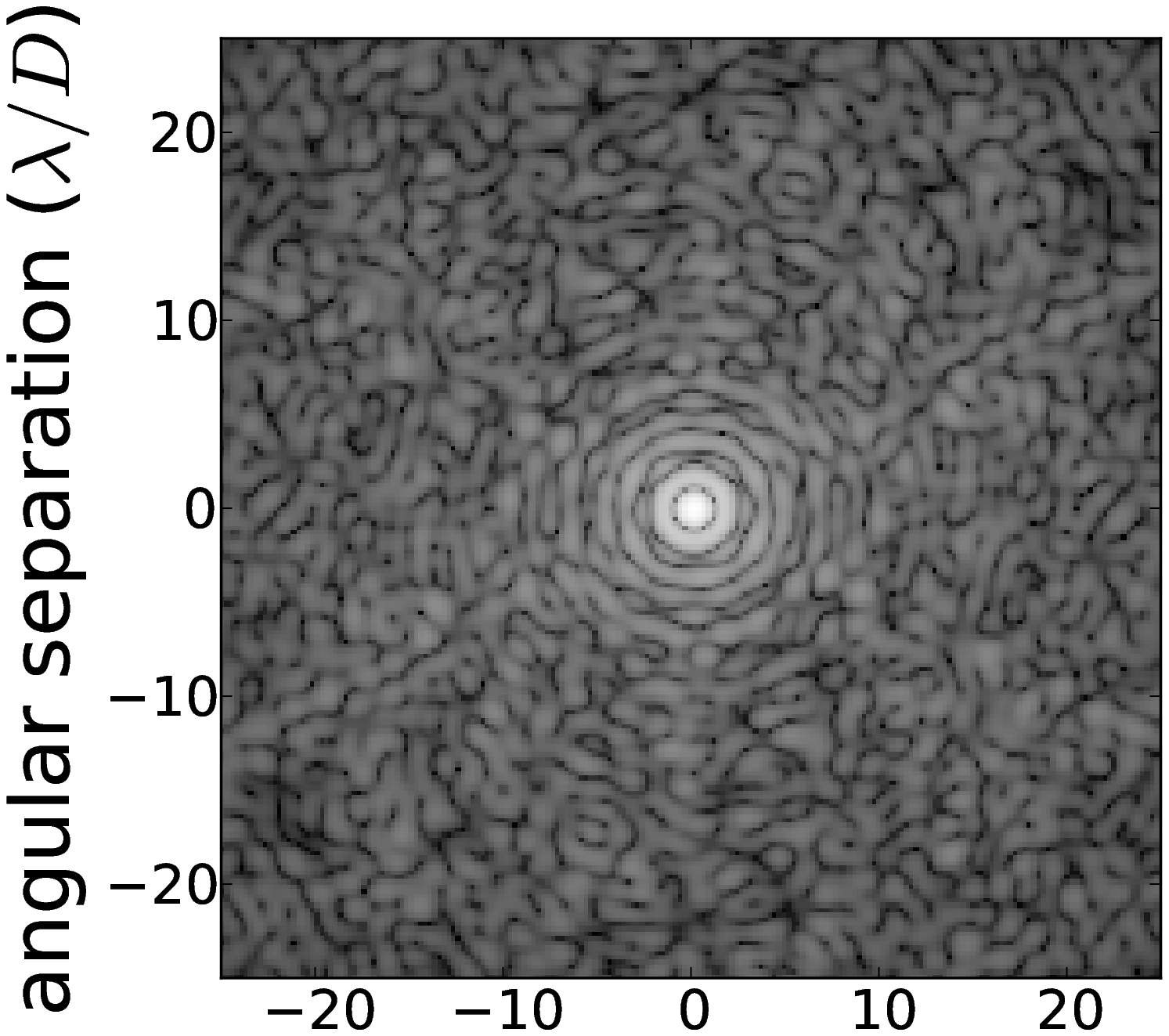} & 
  \includegraphics[width=30mm]{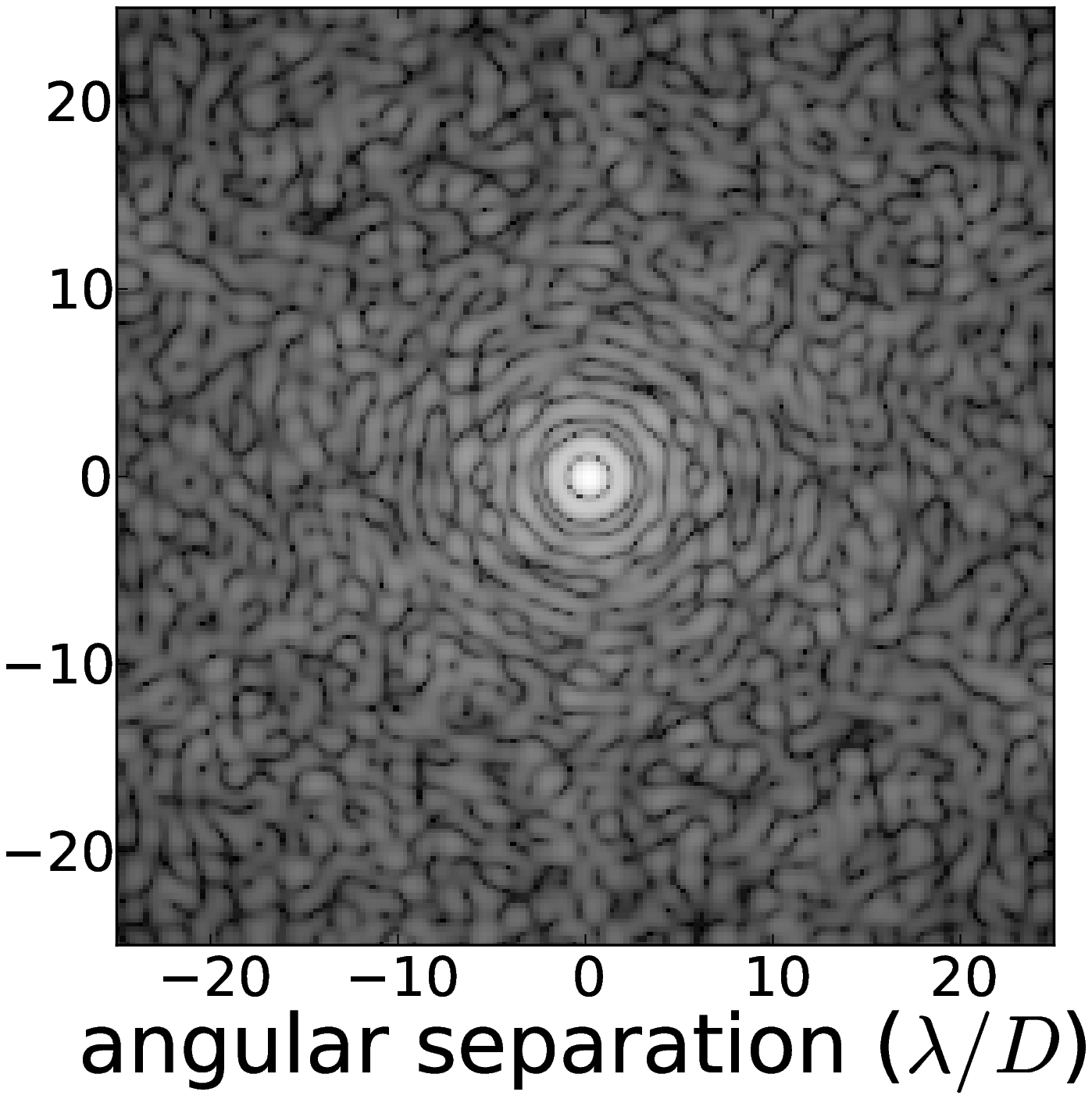} &
  \includegraphics[width=30mm]{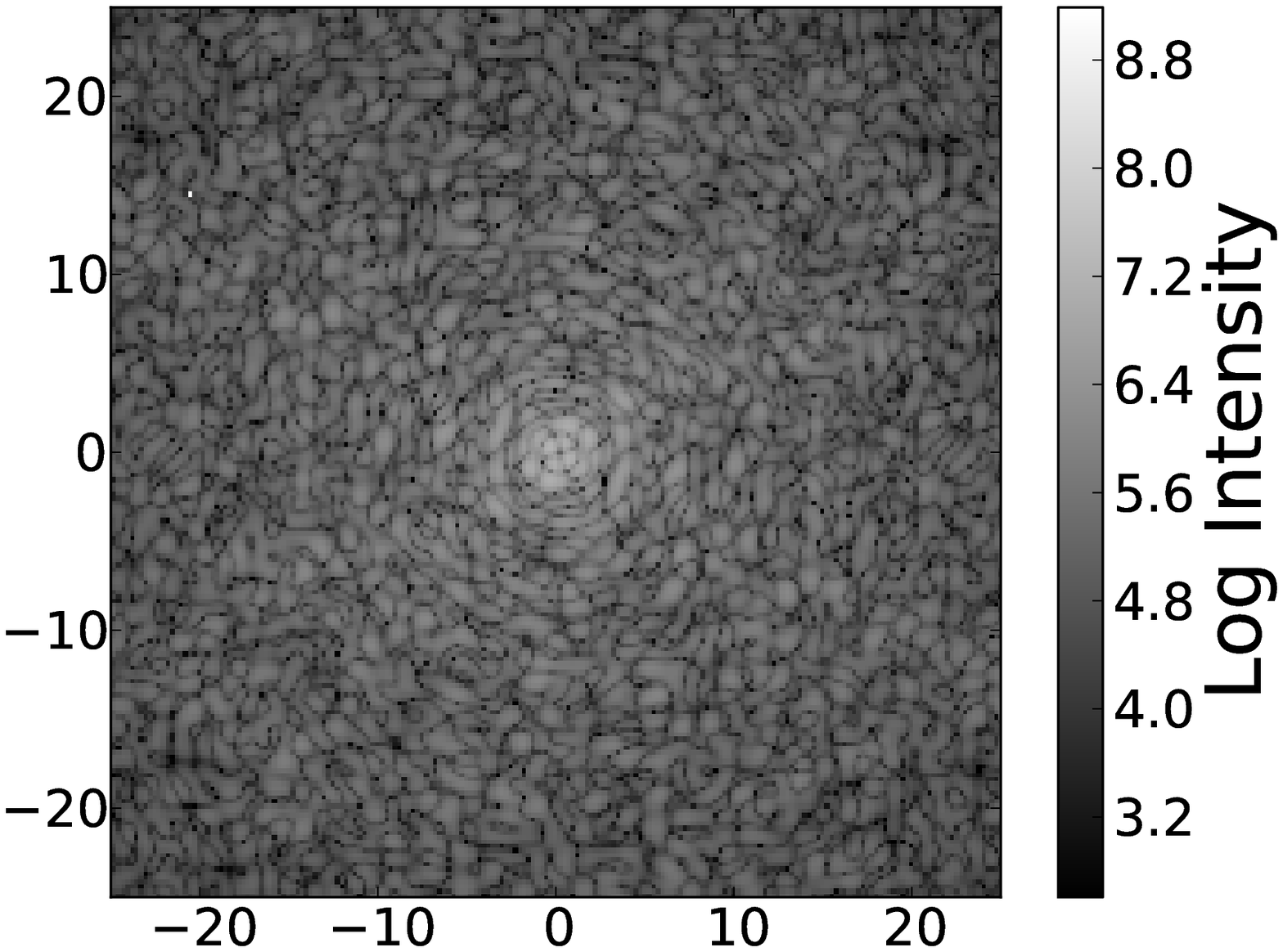} 
   \end{array}$
      \caption{Two individual consecutive images after APM (left and centre). The speckle pattern in each image is different as shown by the modulus of the difference of the images (i.e $|I_1-I_2|$, where $I_1$ and $I_2$ are the intensity patterns of the two frames) in the right panel. In this example we block a random 15\% of the pupil.}
   \label{fig:APM_3f}
\end{figure}

\begin{figure}
   \centering
         \includegraphics[width=60mm]{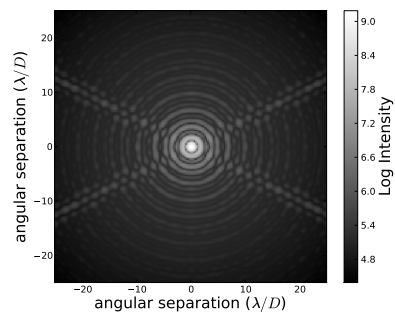} 
      \caption{The sum of 1000 images. In this example we block a random 15\% of the pupil.}
   \label{fig:APM_av}
\end{figure}

\begin{figure}
   \centering
         \includegraphics[width=60mm]{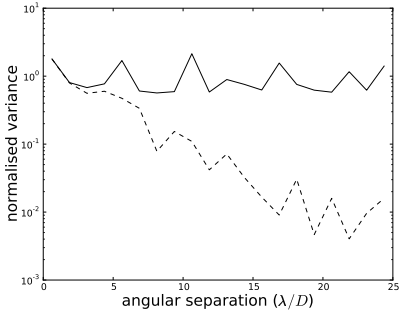} 
      \caption{Azimuthal variance as a function of radius for the original image (black line) and the APM image (dashed line).}
   \label{fig:APM_var}
\end{figure}

In the modified images the quasi-static speckles are no longer quasi-static but are actually very different in every iteration.  Figure~\ref{fig:APM_3f} shows two consecutive focal images and the modulus of the difference. In each individual frame the speckle pattern is completely different. No atmospheric turbulence was included to generate these plots and so the speckles movement is entirely due to the changing pupil mask.
Figure~\ref{fig:APM_av} shows the sum of 1000 frames.
We can define a metric for the magnitude of the quasi-static speckle as the normalised azimuthal variance,
\begin{equation}
\sigma^2(\rho) = \frac{1}{2\pi\rho}\int_0^{2\pi} \frac{|I(\rho,\theta) - \langle I(\rho) \rangle|^2}{\langle I(\rho)\rangle} d\theta,
\end{equation}
where $\langle I(\rho)\rangle$ denotes the expected intensity of the PSF at radius $\rho$. The metric must be normalised by the azimuthal average or the variance will be biased by the radial intensity of the image. Figure~\ref{fig:APM_var} shows the azimuthal variance for the original image and the APM image. We see that the radial intensity variance for the original image is approximately constant for all field angles, the speckle is dominant over the diffraction halo. The azimuthal variance for the APM image continues to reduce with separation. At small angles the diffraction pattern from the central star is intense and the speckles are `pinned' to the first diffraction rings \citep{Bloemhof01}, even with the APM. At larger angles the speckles are more free to move and so we achieve a greater suppression.

When we sum the images over a number of iterations, each with a different pupil pattern, the quasi-static speckles average out to a smooth PSF. However, as we are using square blocking elements Babinet's principle dictates that we can expect the diffraction limited PSF to be a superposition of a square diffraction pattern of the blocking elements and the circular diffraction pattern from the pupil. The diffraction limited PSF in polar co-ordinates is therefore given by,
\begin{multline}
PSF(\rho, \theta) =  \frac{1}{(1-\alpha^2)^2}   \left( \frac{ 2J_1 (\rho D/2)}{ \rho D/2}    -\alpha^2 \frac{2J_1(\alpha\rho D/2)}{\alpha\rho D/2}   \right)^2  \\(1-f) + 
                                2a^2 \mathrm{sinc}\left(\frac{\rho a \sin(\theta)}{2} \right)\mathrm{sinc}\left(\frac{\rho a \cos(\theta)}{2}\right)  f,
\label{eq:diff_theory}
\end{multline}
where $\rho$ is related to the pupil plane parameters by $\rho=2\pi \xi/\lambda f_L$, $\lambda$ is the wavelength of the light and $f_L$ is the effective focal length of the optical system, $\alpha$ is the fractional radius of the central obscuration and $a$ is the length of one side of one of the masking elements.

As the mask elements are smaller than the pupil, the diffraction pattern is broader, increasing the halo around the PSF. Smaller blocking elements will result in a broader diffraction halo and a greater blocking fraction will result in a diffraction halo with a greater fraction of the total energy.

Although the sum of the individual frames does show a reduction in the quasi-static speckles, the disadvantage would be the decreased the signal to noise ratio ({\it SNR}) due to the increased halo intensity and reduced peak intensity. This means that we  would have to integrate for longer to achieve the same {\it SNR}, although the quasi-static speckles will no longer form a fundamental limit -- we are now photon noise limited. 

The theoretical form of the APM PSF is derived in \cite{Osborn10b}. Here we briefly review the theoretical structure of the PSF for the case of random pupil blocking. The PSF, assuming on-axis observations, can be estimated from the modulation transfer function of the residual atmospheric phase aberrations and of the telescope aperture. The PSF is given by,
\begin{equation}
	PSF = \left(\mathcal{F}\left[{\it MTF}_{\rm{AO}}\times {\it MTF}_{\rm{tel}}\right] \right)\times (1-f),
\label{eqn:PSF}
\end{equation}
where  ${\it MTF}_{\rm{AO}}$ is the atmospheric modulation transfer function and $ {\it MTF}_{\rm{tel}}$ is the telescope modulation transfer function. As the ${\it MTF}$ is defined for unit total intensity the PSF must be scaled by $(1-f)$. We can see that the total intensity of the image is reduced by $f$ but the change in peak intensity is not so obvious. It has been shown that with careful pupil blocking the peak intensity can actually be increased \citep{Osborn09}. However, here we choose to randomly block the pupil. In this case we would expect the peak intensity to be reduced. Peak intensity is equal to the integral of the MTF,
\begin{equation}
I_0 = \int_0^\infty ({\it MTF}_{\rm{AO}}\times {\it MTF}_{\rm{tel}} )d\nu (1-f),
\label{eqn:peakint}
\end{equation}
where $\nu$ is the spatial frequency in the focal plane and is related to the separation in the pupil plane, $r$ by $r=\lambda f_L \nu$. The ${\it MTF}$ is related to the phase structure function $D$ by,
\begin{equation}
{\it MTF}_{\rm{AO}}(\nu)=\exp{(-0.5D_{\rm AO}(\nu))},
\end{equation}
and \citep{Rao00}
\begin{equation}
D_{\rm AO}(\nu) = 4\pi \int_0^\infty [1-J_0(\kappa \nu)] \Phi_{AO} \kappa d\kappa.
\end{equation}
Therefore,
\begin{multline}
{\it MTF}_{\rm{AO}}(\nu)=\exp\left(-0.5 \times 4\pi \int_0^\infty [1-J_0(\kappa \nu)] \Phi_{AO} \kappa d\kappa   \right).
\end{multline}
${\it MTF}_{\rm{tel}}$ is given by the autocorrelation function of the aperture function. This will be the product of the pupil function and the mask,
\begin{equation}
W(\xi,\eta,t)=M(\xi,\eta,t)P(\xi,\eta),
\end{equation}
and so,
\begin{equation}
{\it MTF}_{\rm{tel}}(\nu) = \mathcal{F} \left[ |W(\xi,\eta,t)|^{2} \right]
\end{equation}
The mask will act to reduce ${\it MTF}_{\rm{tel}}$ due to the lower fill factor in the pupil (less redundency in Fourier baselines), reducing the peak intensity of the image. Using the above we can write equation~\ref{eqn:PSF} as,
\begin{multline}
	PSF = \left(\mathcal{F}\left[      \exp   \left(-0.5 \times 4\pi \int_0^\infty [1-J_0(\kappa \nu)] \Phi_{AO} \kappa d\kappa   \right) \right.\right. \\
         \left.\left.              \times  \mathcal{F} \left[ |W(\xi,\eta,t)|^{2} \right]  \bigg)    \right] \right)\times (1-f),
\end{multline}
and equation~\ref{eqn:peakint} as,
\begin{multline}
I_0=  \int_0^\infty \bigg( \exp   \left(-0.5 \times 4\pi \int_0^\infty [1-J_0(\kappa \nu)] \Phi_{AO} \kappa d\kappa   \right)  \\
                       \times  \mathcal{F} \left[ |W(\xi,\eta,t)|^{2} \right]  \bigg) d\nu \times (1-f),
\end{multline}
This is solved numerically and for a random blocking of 15\% we find that the peak intensity is reduced by $\sim$27\%. To take a specific example, if we would like to observe a companion at $5\lambda/D$ the normalised halo would be increased by $\sim$16\% in comparison to the non-blocked image.

The exposure time, $t_0$, to achieve a given ${\it SNR}$ assuming we are limited by the photon noise of the halo is given by, 
\begin{equation}
t_0={\it SNR}^2c/SA\epsilon,
\end{equation}
where $S$ is the signal from the planet, $A$ the telescope area, $\epsilon$ is the efficiency of the optical system and $c$ is the contrast ratio between the intensity at the position of the image and the companion signal, $c=n_{\gamma, \rm{h}}(\rho=5\lambda/D)/n_{\gamma, \rm{comp}}(\rho=0)$, where $n_{\gamma, \rm{h}}(\rho=5\lambda/D)$ is the number of photons in the halo at the position, $\rho=5\lambda/D$ and $n_{\gamma, \rm{comp}}(\rho=0)$ is the peak number of photons from the companion.

As the mask will increase the PSF halo and reduce the companion signal the contrast will be increased, requiring longer exposure times to conserve the SNR. For the ratio $t_0/t_{\it APM}$ this simplifies to,
\begin{equation}
t_0/t_{\it APM} = (1-f)c/c_{\it APM},
\end{equation}
where $t_{\it APM}$ and $c_{\it APM}$ is the exposure time and contrast ratio of the APM image respectively. Using the figures found numerically above we see that the contrast is increased by a factor of 1.59 for a companion at $5\lambda/D$. The exposure time to achieve the same {\it SNR} in the noiseless case would be 1.35 times longer than the non-blocked image. This is confirmed by the Monte-Carlo simulation with a value of 1.37. With noise (background, read and shot noise) the value is increased to 1.47. 

\section{Dark speckle imaging}
As we have forced the quasi-static speckle pattern to be dynamic we can use the individual frames to increase the image quality. One suggestion, developed here, would be to use the dark speckle analysis of \cite{Labeyrie95}. 

In each frame the speckle pattern will have minima in intensity. The location of these nulls will vary as the pupil mask changes shape. In the position of the faint companion the probability of seeing a speckle null is considerably lower than anywhere else. 

\begin{figure}
   \centering
    \includegraphics[width=60mm]{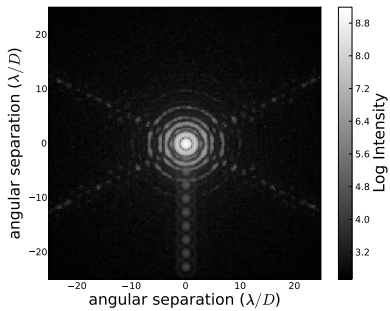} 
   \caption{Dark speckle image. Plotted is the minimum count of each pixel over every iteration of the mask. Eight of the nine companions are now clearly visible. The ninth, the closest, companion is still hidden in the halo of the bright star.}
   \label{fig:APM_DS}
\end{figure}

The proposed technique here is slightly different to that of Labeyrie. Dark speckle imaging is not applicable to long exposures as the atmospheric speckles will average over time, reducing the depth of the nulls. However, here we are not necessarily restricted to very short exposures and hence the read out noise, sky background and AO residuals will mean that the probability of measuring zero photons over an arbitrary time period is low. Instead we record the lowest value of each pixel over all of the iterations of mask configuration. We have chosen to expose 1000 iterations at 3~s exposure time each, resulting in a total observing time of approximately 50 minutes. Figure~\ref{fig:APM_DS} shows the dark speckle image. We see that eight of the companions are now easily visible. The companion closest to the bright star (at 2.5$\lambda/D$) is still hidden in the halo from the bright star.

A disadvantage of this technique is that the intensity at the location of the companion will be the minimum flux measured from the companion over an exposure time of one iteration. This will have the expected value of the intensity of the companion minus the shot noise. For this technique to work a large number of readouts are required, although it should be noted that as we are only recording the minimum value of each pixel we are essentially only including one readout per pixel.  

\begin{figure}
\centering
$\begin{array}{c@{\hspace{-3.5mm}}c}
\includegraphics[width=44mm]{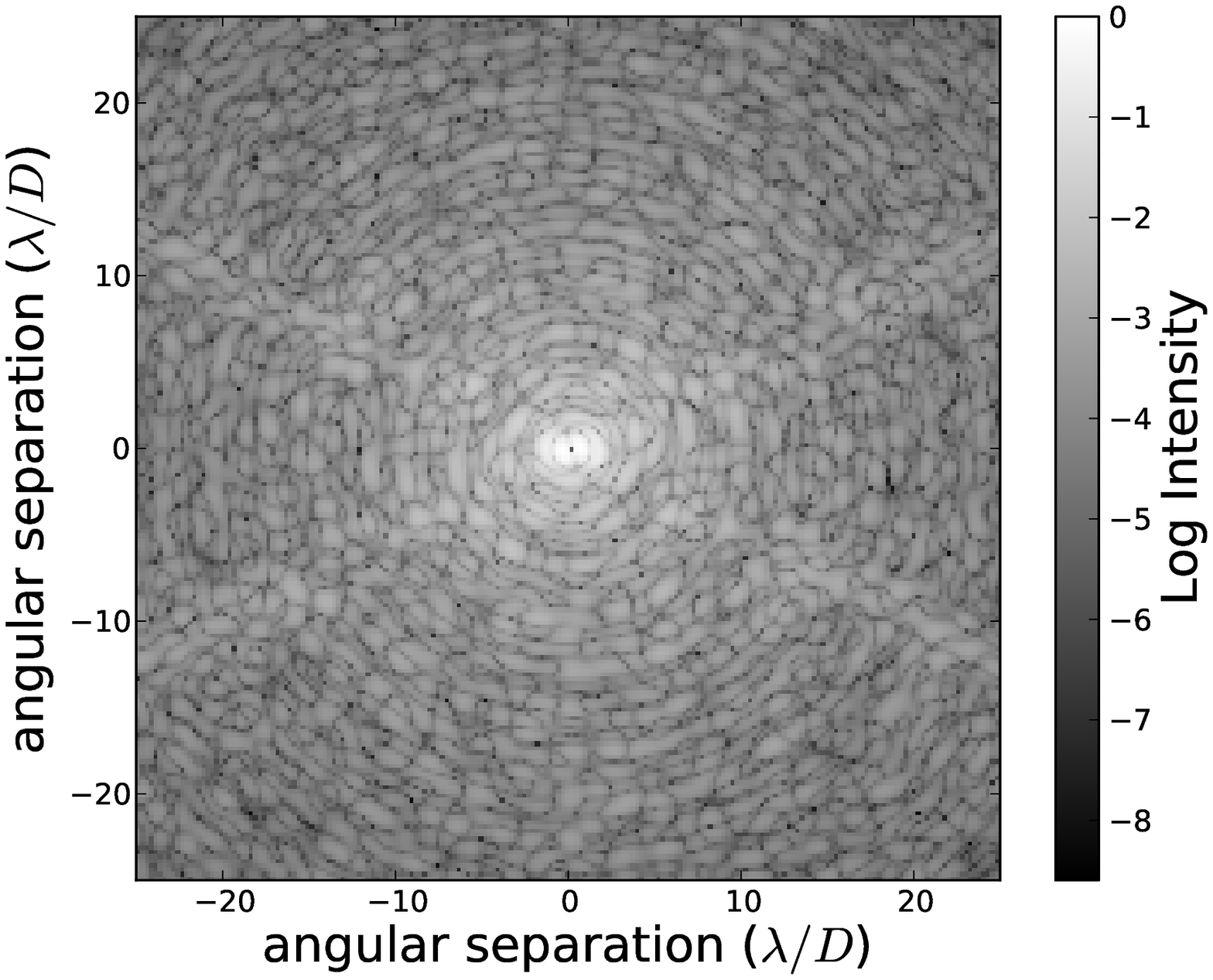} &
 \includegraphics[width=44mm]{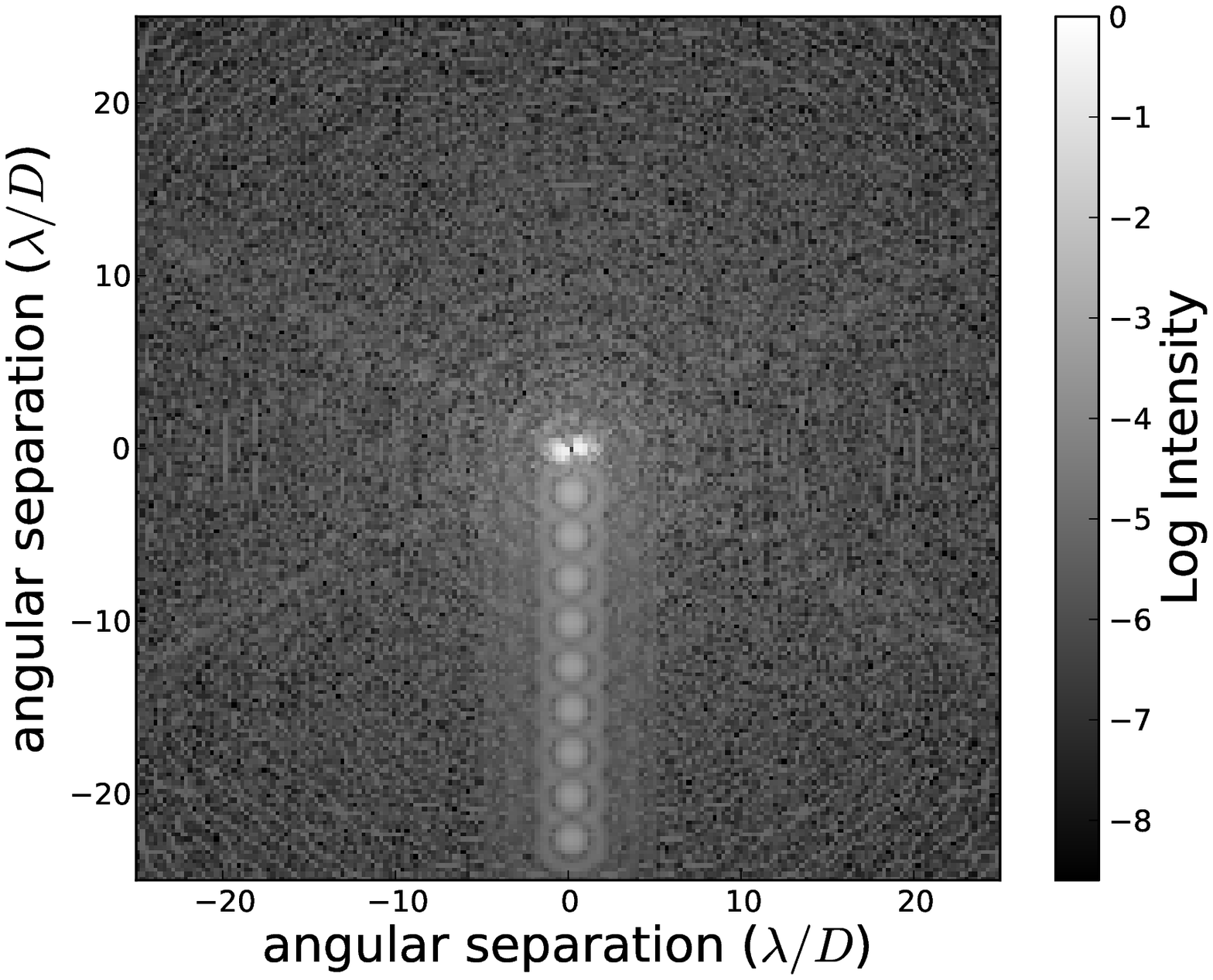} 
 \end{array}$
 \caption{Focal images assuming a perfect coronagraph. The images shown are for the original image (left) and the DS image (right).}
 \label{fig:magicCoron}
\end{figure}

A coronagraph would reject a large fraction of the photons from the parent star and improve the results at small angular separations. For example, if we assume a perfect coronagraph (i.e. one which removes the diffraction limited PSF but not the quasi-static speckles, noise or any diffraction effects due to the mask) then figure~\ref{fig:magicCoron} shows the original image and the DS image. All companions are now visible with very little speckle.


\section{Discussion}
\begin{figure}
\centering
$\begin{array}{c@{\hspace{-3.5mm}}c}
 \includegraphics[width=44mm]{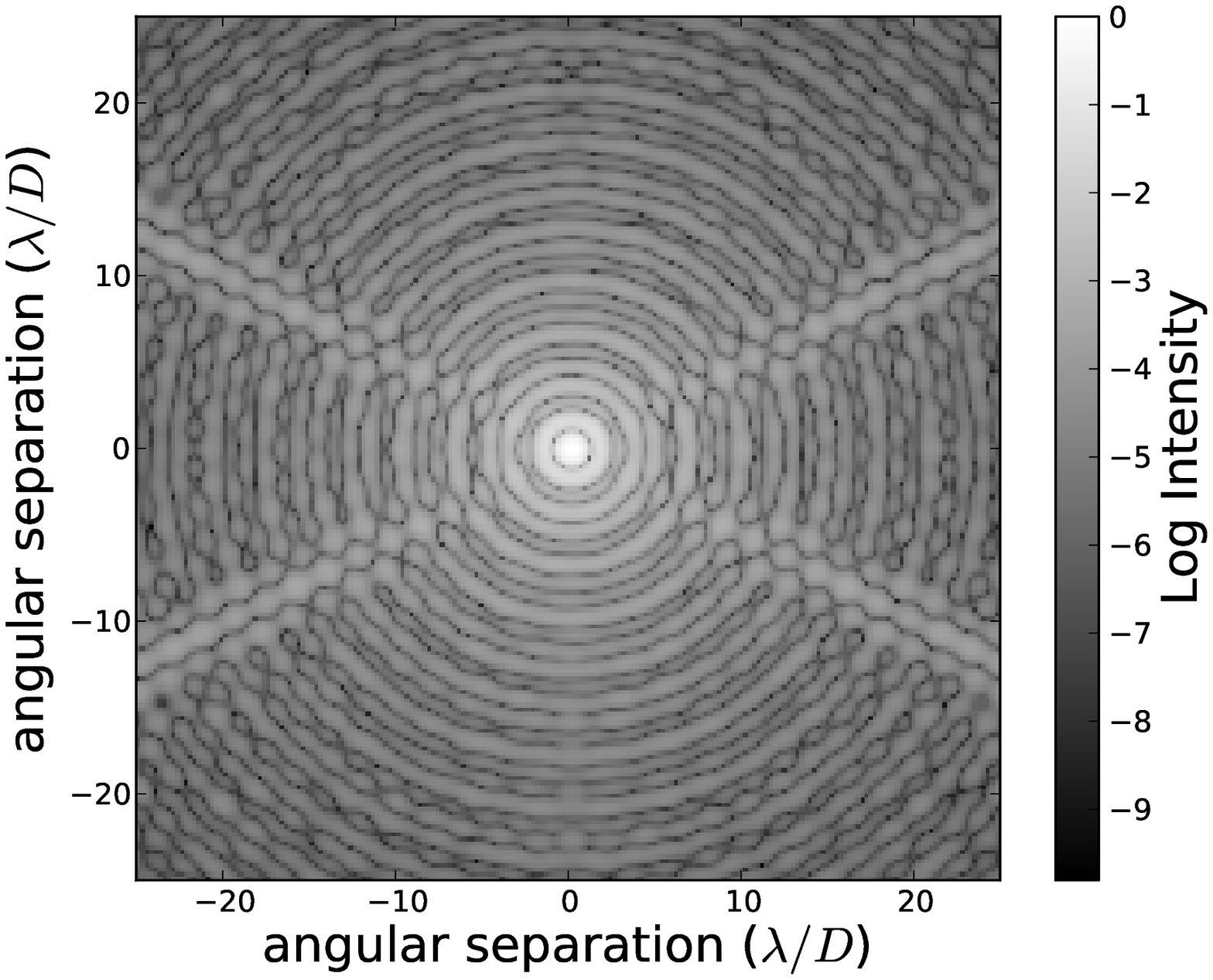} &
 \includegraphics[width=44mm]{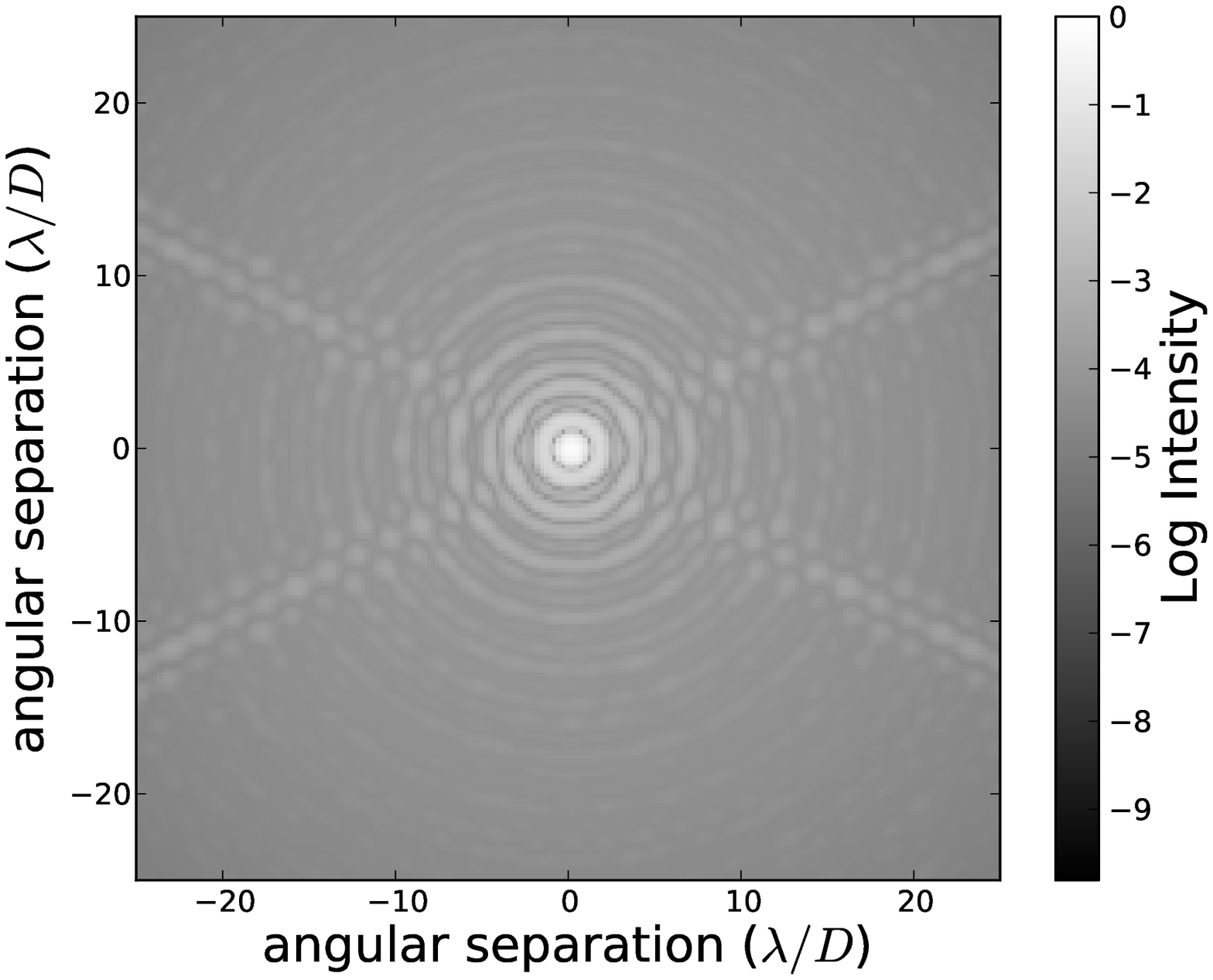} \\
  \mbox{\bf (a)} & \mbox{\bf (b)}\\
   \end{array}$
  \includegraphics[width=44mm]{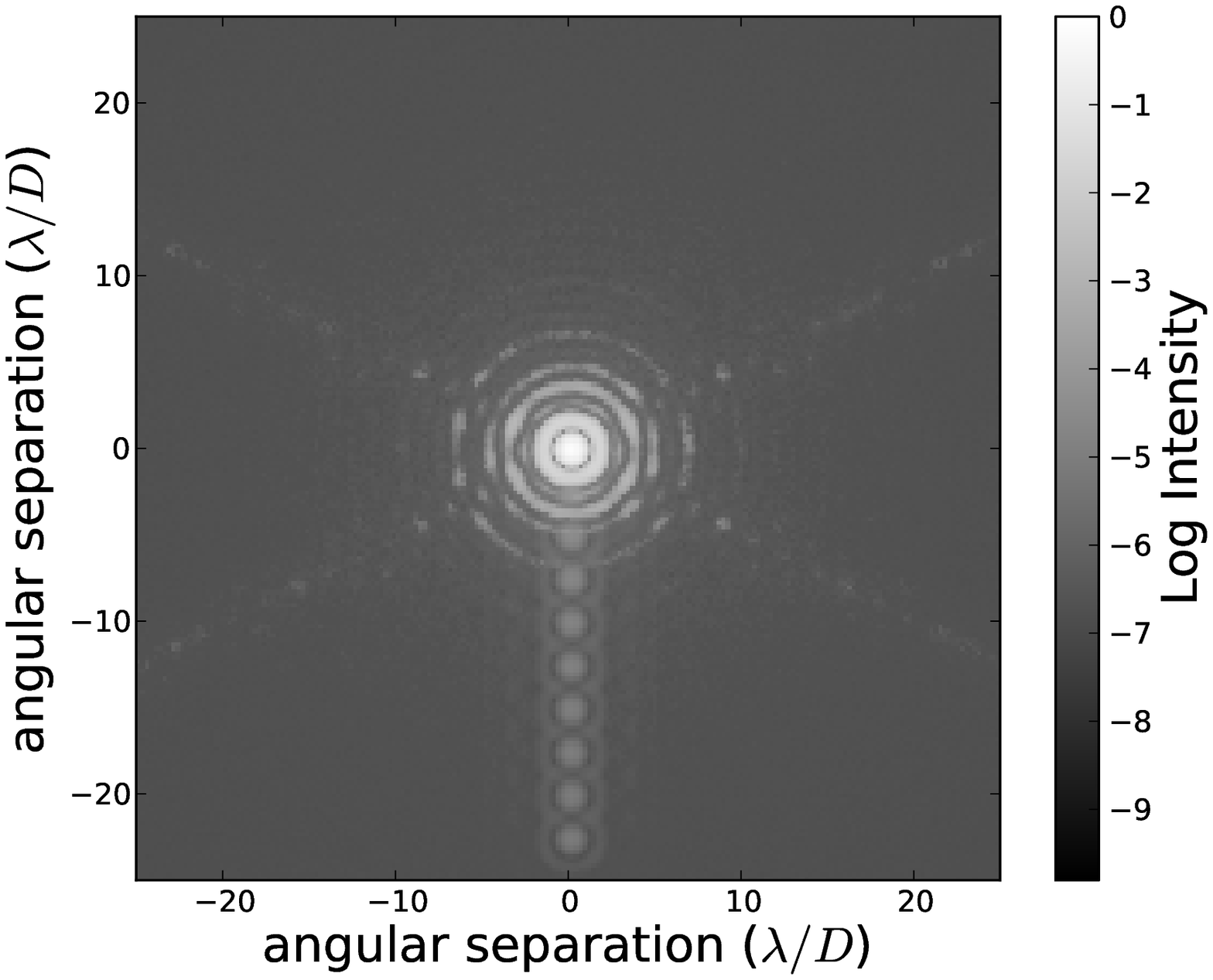}\\
 \mbox{\bf (c)}
 \caption{Focal images including nine faint companions aligned from image centre to the bottom. The images shown are for the case with no manipulation (a), long exposure APM image (b) and APM + DS image (c). All images are now normalised relative to the peak intensity of the original image.}
 \label{fig:final_images}
\end{figure}

\begin{figure}
   \centering
    \includegraphics[width=75mm]{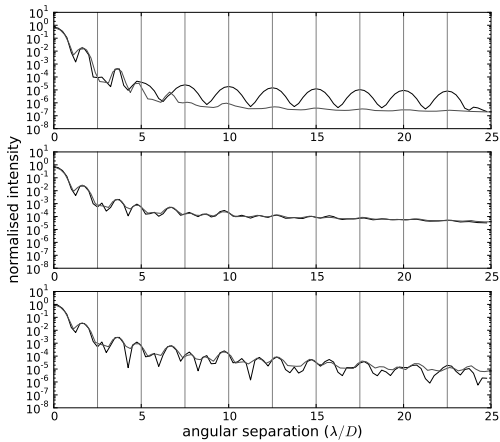} 
   \caption{Radial cut along the direction on the simulated companions (black) and azimuthal average (grey) for the thresholded APM+DS image, long exposure AP image and the original image (from top to bottom). The vertical lines indicate the positions of the companions in the focal plane.}
   \label{fig:radial_cut}
\end{figure}

\begin{figure}
   \centering
    \includegraphics[width=75mm]{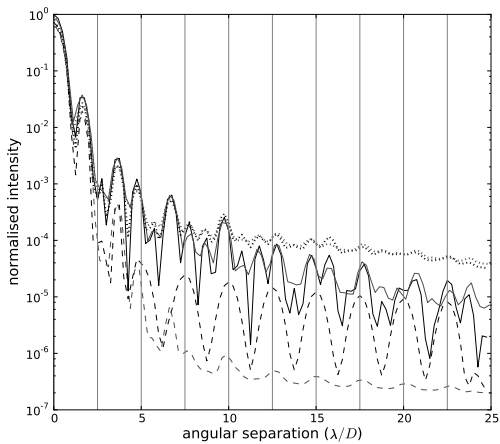} 
   \caption{Radial cut along the direction on the simulated companions (black) and azimuthal average (grey) for the APM+DS image (dashed line), the long exposure APM image (dotted line) and the original image (thin line). The vertical lines indicate the positions of the companions in the focal plane.}
   \label{fig:radial_comp}
\end{figure}

For comparison, figure~\ref{fig:final_images} shows the final images including the nine faint companions for the case of no manipulation, long exposure APM and APM with DS image analysis.
We can examine radial cuts of the final images to compare performance. In figure~\ref{fig:radial_cut} we show a radial cut in the direction of the companions and the azimuthal average for the final APM+DS image (top), long exposure APM image (middle) and the original image (bottom). The vertical lines indicate the position of the faint companion. 
Figure~\ref{fig:radial_comp} shows the radial profile of the original image the long exposure APM image and the APM+DS image on the same axis. 

We can see that the effect of the companions is visible in the original and long exposure images, but only the APM+DS method has been able to isolate all but the closest companion successfully in this case. At separations greater than approximately $5\lambda/D$ the contrast between the PSF peak and the halo is increased by the DS method by a factor of approximately 100, corresponding to 5 magnitudes. It is clear that not only have the speckles been suppressed but also the bright star diffraction pattern is also suppressed. This is because the speckles nulls are lower than the average intensity of the PSF halo for any given location. We are now able to detect companions 11 magnitudes fainter than the star at separations of $5\lambda/D$ and 18 magnitudes fainter at $22.5\lambda/D$. The inner working angle would be improved with the use of a coronagraph or other PSF subtraction methods which is not included here.

The performance of the technique is dependant on the parameters selected and on the properties of the system we wish to observe. Detection is modified if we increase the fraction that the APM blocks, the number of iterations or the Strehl ratio of the AO system, of which, only the blocked fraction of the mask and the number of iterations are controlled by the observer. If we increase the fraction which is blocked then we generate more speckle in the focal plane, which means quicker averaging and more nulls.  However, it will also mean reducing the throughput of the system, therefore this will depend on the system to be observed. In the simulation we blocked 15\% of the pupil. The number of iterations of the mask is also important. The more iterations the greater number of configurations of the mask are used. This means more averaging in the long exposure APM image and means more nulls are measured in the DS image, increasing detection. The number of iterations is a balance between the amount of time available for an observation and the exposure time of each iteration. In order to receive a significant number of photons from the companion it may be necessary to increase the exposure time, this will limit the possible number of iterations. In the simulations we assumed 1000 iteration of 3 seconds each. It should be noted that there is no limit to the exposure time of each iteration.


\section{Conclusions}

Atmosphere induced phase aberrations (with or without AO) causes a halo of speckles to form around the PSF. This source of noise will average over time and, given enough time, faint companions can be observed above this halo.  Current high contrast imaging is limited by quasi-static speckles caused by the optics and structure of the telescope. This is because these quasi-static speckles do not average. They appear in the image as potential false-positive candidates for faint companions and are difficult to distinguish. Here we have presented a technique to turn these static speckles into dynamic speckles. This means that over time these static speckles will also average into a broad halo. The shape and magnitude of this halo will depend upon the geometry of each mask element and the fraction of the pupil that is blocked. This means that over time the PSF will converge to a smooth form allowing the companions to be seen above. However, as the APM adds more energy to the halo and the throughput of the system is reduced by a fraction equal to the fraction of the pupil which is blocked, the {\it SNR} is reduced. We would need to observe for a longer period of time to collect enough photons from the companion to be seen above the halo. For an example case of a companion at $5\lambda/D$ we would need to observe for 1.35 times longer (1.47 defined by the simulation including noise) to achieve the same {\it SNR}. In this case the we do not need to expose the CCD between each mask state and so the mask frequency can be high allowing many mask states for an arbitrarily long exposure.

In addition to the smoother PSF we can also use other conventional image manipulation techniques which are normally used for the dynamic speckling caused by atmospheric turbulence. Here we show the effect of implementing an adaptation of Labeyrie's dark speckle analysis. This is a simple technique where we record the minimum value of each pixel over all of the iterations of the mask. 

Using this technique in a Monte-Carlo simulation we can detect faint companions with a higher magnitude difference from the central star. We find that at separations greater than a few $\lambda/D$ the PSF halo count is reduced by a factor of approximately 100, corresponding to a contrast increase of 5 magnitudes. 

Due to the APM the diffraction rings from the bright star also move, this means that the diffraction pattern from the bright star is also suppressed by the APM when used with the DS imaging analysis. At small inner working angles there is still some confusion between the companions and that of the parent star, due to the pinned speckles in the bright diffraction rings. However, we are now able to detect companions 11 magnitudes fainter than the star at separations of $5\lambda/D$ and up to 18 magnitudes fainter at $22.5\lambda/D$. The inner working angle would be improved with the use of a coronagraph or other PSF subtraction methods which is not included here. In order to perform the DS analysis short exposure images will have to be recorded. The length of the short exposures is arbitrary but the optimal will depend on the number of iterations required during the exposure and the magnitude of the target.

The noise attenuation found here is comparable with the attenuation from the High-Order Testbench (HOT) with ADI \citep{Martinez12}. The advantage of this technique over ADI is that it does not rely on PSF subtraction and is therefore insensitive to any changes in the quasi-static pattern. It has the advantage over PDI and SSDI that it also insensitive to the companion properties, i.e. it does not require the companion to have a specific emission spectrum or the light to be reflected and hence partially polarised. The disadvantage is the reduced throughput.

APM would benefit from collaboration with other techniques. Currently the inner working angle is limited by the diffraction pattern of the primary star. A coronagraph would reject light from the central star resulting in even higher achievable contrast ratios. Also, no additional image manipulation was performed. No attempt at PSF subtraction was made. The dynamic speckles will average into the predictable smooth long exposure PSF, which could be subtracted. Combination with other techniques would improve the noise attenuation and reduce the probability of false-positives.

It is also important to note that by including a high order APM, in addition to quasi-static speckle removal, it would also be possible to reduce the residual WFE, emulate a non-redundant (or partially redundant) aperture mask and a binary shaped pupil plane coronagraph, all of which would be completely and easily re-configurable.

\section*{Acknowledgements}
I would like to thank Gordon Love and Andres Guesalaga for their helpful comments. The author received a postdoctoral fellowship from the School of Engineering at Pontificia Universidad Cat\'{o}lica de Chile as well as from the European Southern Observatory and the Government of Chile.

\bibliographystyle{mn2e}

{}

\end{document}